\batchmode
\documentclass[twocolumn,prb,floatfix]{revtex4}% Physical Review B RC

\usepackage{graphicx}
\usepackage{dcolumn}
\usepackage{amsmath}

\begin{document}

\title{Ultrafast spectroscopy of propagating coherent acoustic phonons
in GaN/InGaN heterostructures}

\author{Rongliang Liu,
Chang Sub Kim\footnote{Permanent address: Department of Physics,
Chonnam National University, Kwangju 500-757, Korea},
G. D. Sanders, and C. J. Stanton}
\affiliation{Department of Physics, University of Florida, Box 118440\\
Gainesville, Florida 32611-8440}

\author{J. S. Yahng, Y. D. Jho, K. J. Yee, E. Oh, and D. S. Kim}
\affiliation{Department of Physics, Seoul National University, Seoul
151-742, Korea}

\begin{abstract}

We show that large amplitude, coherent acoustic phonon wavepackets can
be generated and detected in In$_x$Ga$_{1-x}$N/GaN epilayers and
heterostructures in femtosecond pump-probe differential reflectivity
experiments.  The amplitude of the coherent phonon increases with
increasing Indium fraction $x$ and unlike other coherent phonon
oscillations, both \textit{amplitude} and \textit{period} are strong
functions of the laser probe energy.  The amplitude of the oscillation
is substantially and almost instantaneously reduced when the
wavepacket reaches a GaN-sapphire interface below the surface
indicating that the phonon wavepackets are useful for imaging below
the surface.  A theoretical model is proposed which fits the
experiments well and helps to deduce the strength of the phonon
wavepackets.  Our model shows that localized coherent phonon
wavepackets are generated by the femtosecond pump laser in the
epilayer near the surface. The wavepackets then propagate through a
GaN layer changing the local index of refraction, primarily through
the Franz-Keldysh effect, and as a result, modulate the reflectivity
of the probe beam. Our model correctly predicts the experimental
dependence on probe-wavelength as well as epilayer thickness.

\end{abstract}
\date{\today}

\maketitle

%%%%%%%%%%%%%%%%%%%%%%%%%%%%%%%%%%%%%%%%%%%%%%%%%%%%%%%%%%%%%%%%%%%%
\section{Introduction}
%%%%%%%%%%%%%%%%%%%%%%%%%%%%%%%%%%%%%%%%%%%%%%%%%%%%%%%%%%%%%%%%%%%%

Heterostructures of GaN and InGaN are important materials owing to
their applications to blue laser diodes and high power
electronics.\cite{Nak2000} Strong coherent acoustic phonon
oscillations have recently been detected in InGaN/GaN multiple quantum
wells.\cite{Sun00179,Ozg0185} These phonon oscillations were much
stronger than folded acoustic phonon oscillations observed in other
semiconductor superlattices.\cite{Yam94740,Bar991044,Miz998262}

InGaN/GaN heterostructures are highly strained at high In
concentrations giving rise to large built-in piezoelectric
fields,\cite{Lef011252,Tak981691,Im98R9435,Chi982006} and the large
strength of the coherent acoustic phonon oscillations was attributed
to the large strain and piezoelectric fields.\cite{Sun00179}

In this paper, we report the generation of strong localized coherent
phonon wavepackets in strained layer In$_x$Ga$_{1-x}$N/GaN epilayers
and heterostructures grown on GaN and Sapphire
substrates.\cite{Yah024723} By focusing high repetition rate,
frequency-doubled femtosecond Ti:Sapphire laser pulses onto strongly
strained InGaN/GaN heterostructures, we can, through the
carrier-phonon interaction, generate coherent phonon wavepackets which
are initially localized near the epilayer/surface but then propagate
away from the surface/epilayer and through a GaN layer. As the
wavepackets propagate, they modulate the local index of refraction and
can be observed in the time-dependent differential reflectivity of the
probe pulse.  There is a sudden drop in the amplitude of the
reflectivity oscillation of the probe pulse when the phonon wavepacket
reaches a GaN-sapphire interface below the surface.  Theoretical
calculations as well as experimental evidences support this picture;
the sudden drop of amplitude when the wave encounters the GaN-sapphire
interface cannot be explained if the wavepacket had large spatial
extent. When the wavepacket encounters the GaN-sapphire interface,
part of the wave gets reflected while most of it gets transmitted into
the sapphire substrate, depending on the interface properties and the
excess energy of the exciting photons. This experiment illustrates a
non-destructive way of generating high pressure tensile waves in
strained heterostructures and using them to probe semiconductor
structure below the surface of the sample.  Since the strength of this
non-destructive wave is determined by the strain between GaN and
InGaN, it is likely that even stronger coherent phonons can be
generated in InGaN/GaN digital alloys grown on a GaN substrate.

%%%%%%%%%%%%%%%%%%%%%%%%%%%%%%%%%%%%%%%%%%%%%%%%%%%%%%%%%%%%%%%%%%%%
\section{Experimental Results}
%%%%%%%%%%%%%%%%%%%%%%%%%%%%%%%%%%%%%%%%%%%%%%%%%%%%%%%%%%%%%%%%%%%%

%...................................................................
%                          Figure 1
%...................................................................
\begin{figure*}[t]
\begin{center}
\includegraphics[width=6.8in]{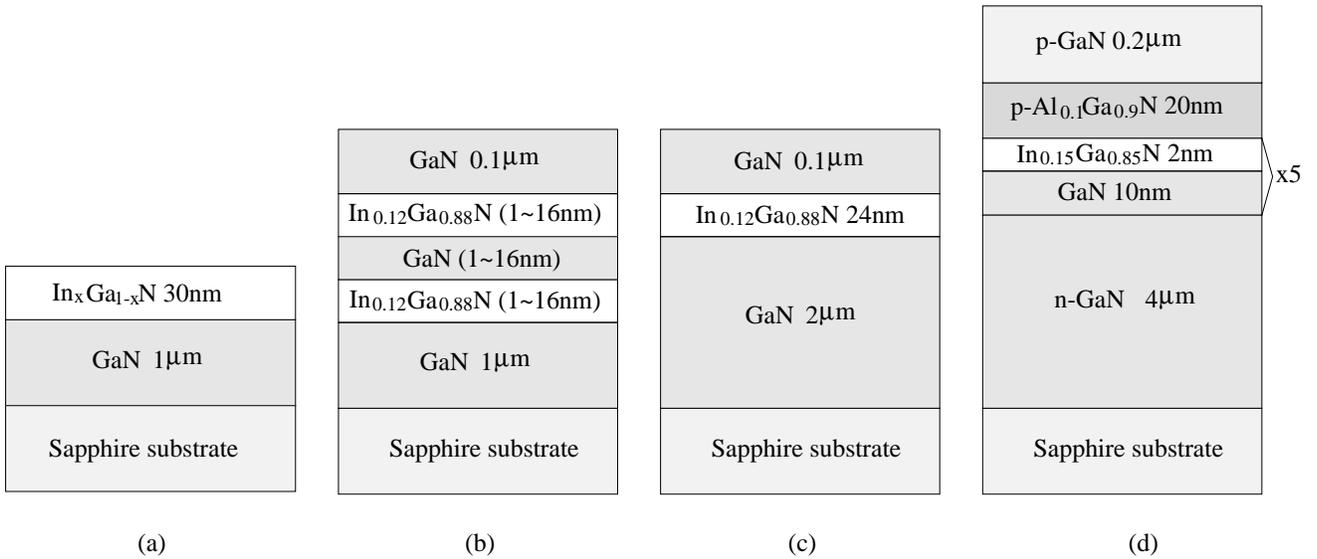}
\end{center}
\caption{ The simple diagram of the sample structures used in
these experiments. (a)InGaN epilayer (type I), (b)InGaN/GaN double
quantum wells (type II), (c)GaN single quantum well (type III),
(d)InGaN/GaN blue light-emitting diode structure (type IV).}
\label{fsamples}
\end{figure*}
%...................................................................

In the experiments, frequency-doubled pulses of mode-locked
Ti:sapphire lasers are used to perform reflective pump-probe
measurements on four different sample types which are shown in
Fig.~\ref{fsamples}): Type (I) InGaN epilayers; Type (II) InGaN/GaN
double quantum wells (DQWs); Type (III) InGaN/GaN single quantum well
(SQW); and Type (IV) InGaN/GaN light-emitting diode (LED)
structures. The peak pump power is estimated to be 400~MW/cm$^2$,
corresponding to a carrier density of 10$^{19}$~cm$^{-3}$ and the
doubled pulse width is 250~fs. All samples were grown on a $c$-plane
sapphire substrate by metal organic chemical vapor deposition.  The
InGaN epilayers shown in Fig.~\ref{fsamples}(a) consist of 1~$\mu$m
GaN grown on a sapphire substrate and capped with 30~nm of
In$_x$Ga$_{1-x}$N with In composition, $x$, varying from 0.04 to
0.12. The DQW sample shown in Fig.~\ref{fsamples}(b) consists of GaN
(1~$\mu$m), double quantum wells of In$_{0.12}$Ga$_{0.88}$N (1--16 nm),
a barrier of GaN (1--16 nm), and a GaN cap layer (0.1~$\mu$m). The
SQW sample in Fig.~\ref{fsamples}(c) consists of GaN (2~$\mu$m),
an In$_{0.12}$Ga$_{0.88}$N well (24~nm), and a GaN cap layer
(0.1~$\mu$m).  The
blue LED structure shown in Fig.~\ref{fsamples}(d) consists of $n$-GaN
(4~$\mu$m), 5 quantum wells of In$_{0.15}$Ga$_{0.85}$N (2~nm) and 4
barriers of GaN (10~nm), $p$-Al$_{0.1}$Ga$_{0.9}$N (20~nm), and
$p$-GaN (0.2~$\mu$m).

%...................................................................
%                            figure 2
%...................................................................
\begin{figure}[tbp]
\begin{center}
\includegraphics[width=3.4in]{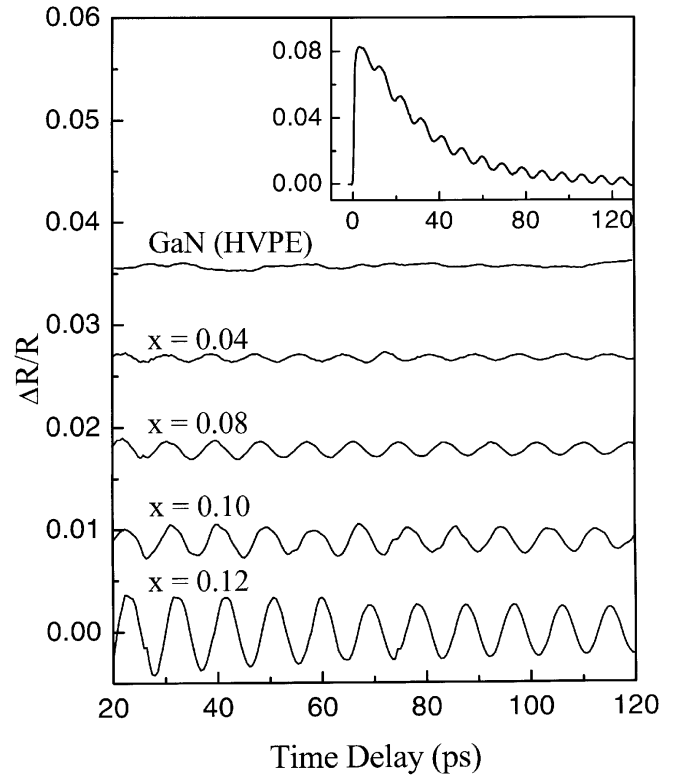}
\end{center}
\caption{The oscillatory component of the differential reflection
pump-probe data for the In$_x$Ga$_{1-x}$N epilayers with various In
composition ($x$=0.04, 0.08, 0.10, and 0.12). For comparison,
differential reflection in a pure GaN HPVE grown sample is shown.
The reflection signal prior to the background subtraction for
$x=0.12$ is shown in the inset.}
\label{fxrtx}
\end{figure}
%..................................................................

Differential reflection pump-probe measurements for the
In$_x$Ga$_{1-x}$N epilayers are shown in Fig.~\ref{fsamples}(a).
Figure~\ref{fxrtx} shows the oscillatory component of the measured
probe differential reflectivity for the InGaN epilayers (type I) with
various In concentrations, x. For comparison purposes, we performed
differential reflectivity measurements on a pure GaN HVPE grown sample
in order to show that no differential reflectivity oscillations are
present in the absence of strain and an epilayer.  The energy of the
pump laser was varied between 3.22 and 3.35~eV, to keep the excess
carrier energy above the InGaN band gap but below the GaN band gap. We
note that if the laser energy was below the InGaN band gap, no signal
was detected. \textit{Therefore, carrier generation is essential to
observing the oscillations, unlike a recent coherent optical phonon
experiment in GaN}\cite{Yee02105501}. The inset shows the pump-probe
signal prior to the background subtraction for $x=0.12$. The
background results from the relaxation of the photoexcited electrons
and holes. The oscillations are quite large, on the order of
$10^{-2}$--$10^{-3}$ and the period is 8--9 ps, \textit{independent}
of the In composition but \textit{dependent} on the probe photon
energy. The amplitude of the oscillation is approximately proportional
to the In concentration indicating that the strain at the InGaN/GaN
interface is important.  The observed period is approximately
$\tau=\lambda/2C_\text{s}n$, where $\lambda$ is the probe beam
wavelength, $C_\text{s}$ the longitudinal acoustic sound velocity, and
$n$ the refractive index of GaN \cite{Tho864129}.

%..................................................................
%                            figure 3
%..................................................................
\begin{figure}[tbp]
\begin{center}
\includegraphics[width=3.4in]{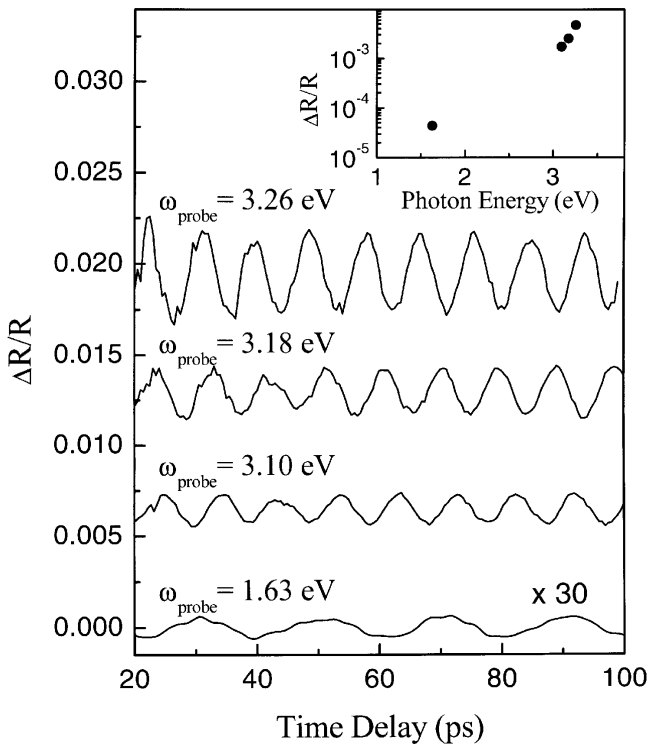}
\end{center}
\caption {The oscillation traces of a SQW (III) at different
probe energies. The pump energy is centered at 3.26~eV. The bottom
curve has been magnified 30 times. The inset shows the oscillation
amplitude as a function of probe photon energy on a logarithmic
scale.}
\label{fxrtw}
\end{figure}
%..................................................................

We performed two-color pump-probe experiments for a type III InGaN SQW
sample as shown in Fig.~\ref{fsamples}(b).  Fig.~\ref{fxrtw} shows the
differential reflectivity oscillations for different probe energies.
Note that the period of the oscillation changes
\textit{and is proportional to the probe wavelength}. In addition, the
amplitude of the differential reflectivity oscillation decreases as
the detuning (with respect to the pump) becomes larger. The
inset shows the oscillation amplitude as a function of the probe
energy in a logarithmic scale and there is an $\sim$2-order-of
magnitude decrease in differential reflectivity when the probe energy
changes from 3.26~eV to 1.63~eV.

%..................................................................
%                              figure 4
%..................................................................
\begin{figure}[tbp]
\begin{center}
\includegraphics[width=3.4in]{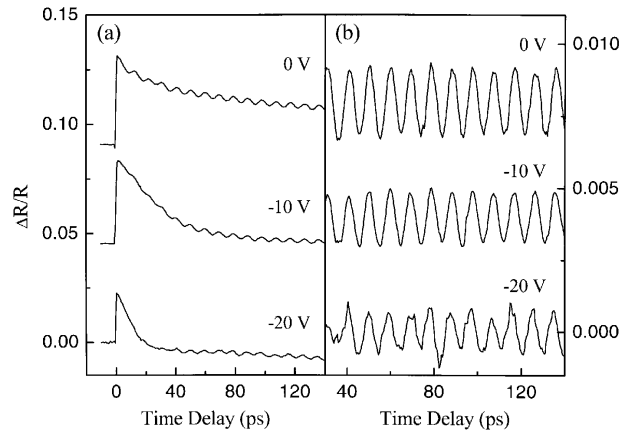}
\end{center}
\caption {(a)Pump-probe differential reflectivity for the blue LED
structure (type IV) varying the external bias at a pump energy of
3.17 eV.  The decay time of the background signal is drastically reduced
as the bias increases. (b) The oscillatory amplitude does not change
much with bias.}
\label{led2}
\end{figure}
%..................................................................

Interestingly, Fig.~\ref{led2} shows that the amplitude of the
oscillatory component of the differential reflectivity is independent
of the bias voltage, even though the carrier lifetime changes
dramatically with voltage bias. Fig.~\ref{led2} shows the bias
dependent acoustic phonon differential reflectivity oscillations in a
type IV blue LED structure (see Fig.~\ref{fsamples}(d)) at a pump
energy of 3.17~eV. The lifetime of the background signal drastically
decreases as the bias increases as shown in Fig.~\ref{led2}(a). This
is due to the carrier recombination time and the decrease in the
tunneling escape time in the strong external bias
regime\cite{Jho011130}. On the other hand, the amplitude and frequency
of the oscillatory component of the differential reflectivity doesn't
change much with bias voltage [Fig.~\ref{led2}(b)]. Since the observed
reflectivity oscillation is independent of the carrier lifetime for
lifetimes as short as 1 ps due to ultrafast tunneling [bottom curve of
Fig.~\ref{led2}(a)], it implies that once the source that modulates
the experimentally observed reflectivity is launched by the
sub-picosecond generation of carriers, the remaining carriers do
little to affect the source. This suggests that the reflectivity
oscillation is due to the strain pulse which is generated at short
times once the pump excites the carriers and modulates the lattice
constant.

%..................................................................
%                           figure 5
%..................................................................
\begin{figure}[tbp]
\begin{center}
\includegraphics[width=3.4in]{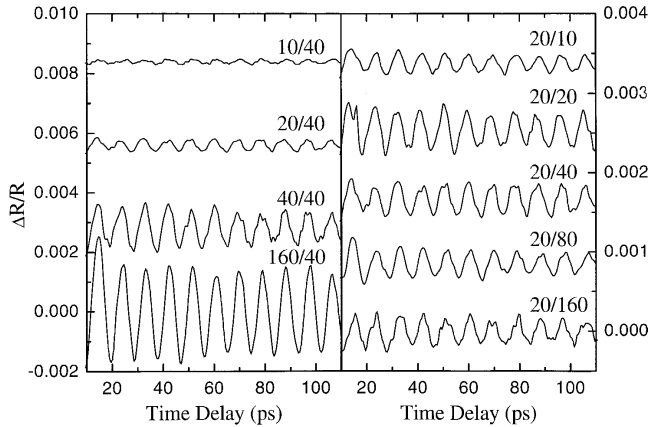}
\end{center}
\caption {The oscillatory component of the pump-probe differential 
transmission traces of DQW's (II) at 3.22 eV. The
left figure shows the well width dependence and the right figure
shows the barrier width dependence.}
\label{dqw3}
\end{figure}
%..................................................................

Figure \ref{dqw3} shows the well- and barrier-dependent acoustic
phonon differential reflectivity oscillations in the type II DQW
samples (see Fig.~\ref{fsamples}(b)) at a pump energy of 3.22 eV. The
amplitude increases as the well width increases. However, the
oscillation amplitude of the differential reflectivity doesn't change
much with the barrier width. This means that the generation of the
acoustic phonons is due to the InGaN well and not the GaN
barrier.  This also verifies that the oscillation is due to the
strain in InGaN layer.

%...................................................................
%                           figure 6
%...................................................................
\begin{figure}[tbp]
\begin{center}
\includegraphics[width=3.4in]{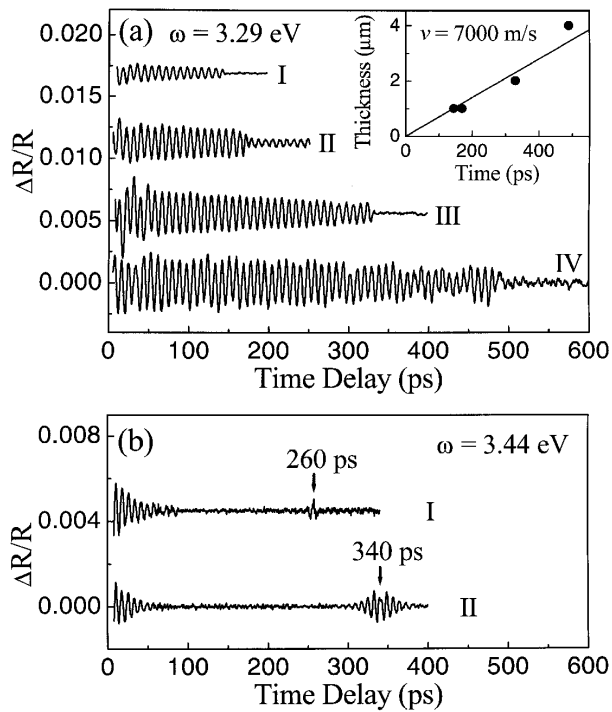}
\end{center}
\caption { (a)The long time-scale oscillation traces of an
epilayer (I), a DQW (II), a SQW (III), and blue LED structure
(IV). The inset shows the GaN thickness between the sapphire
substrate and InGaN active layer of the sample as a function of
the die out time of the oscillations. The solid line indicates
that the velocity of the wavepacket in the GaN medium is about
7000 m/s. (b) The oscillation traces of epilayer and DQW [top two
curves in (a)]at 3.44~eV which corresponds to a probe laser energy
above the band gap of GaN}
\label{fxrtw2}
\end{figure}
%...................................................................

Interesting results are seen in the long time behavior of the
reflectivity oscillations shown in Fig.~\ref{fxrtw2}(a). The long time
scale reflectivity oscillation is plotted for the epilayer (I), DQW
(II), SQW (III), and the blue LED structure (IV) at 3.29 eV (below the
GaN band gap). Astonishingly, the oscillation amplitude abruptly
decreases within one cycle of an oscillation at a critical time which
appears to scale with the thickness of the GaN layer in each
sample. In addition, the slope of the GaN thickness vs. the critical
time is very close to the known value of the sound velocity in GaN
[inset of Fig.~\ref{fxrtw2}(a)]
\cite{Sun00179}. Fig.~\ref{fxrtw2}(b) shows results when the probe
laser energy is changed to 3.44 eV which is above the GaN band
gap. Then the laser probe is sensitive to coherent phonon oscillations
only within an absorption depth of the surface. We see that the
amplitude of the reflectivity oscillation exponentially decays with a
decay time of 24.2 ps corresponding to a penetration depth in GaN of
about 0.17 micron (=24.2~ps$\times$7000~m/s). The oscillation
reappears at 260~ps for epilayer (I) and 340~ps for DQW (II). This is
twice the critical time for the oscillations to disappear when the
photon energy is 3.29~eV. This further shows that the probe pulse is
sensitive to the coherent acoustic wave. The "echo" in the probe
signal results from the partial reflection of the coherent phonon off
the GaN/sapphire interface.

%%%%%%%%%%%%%%%%%%%%%%%%%%%%%%%%%%%%%%%%%%%%%%%%%%%%%%%%%%%%%%%%%%%%%%
\section{Theory}
%%%%%%%%%%%%%%%%%%%%%%%%%%%%%%%%%%%%%%%%%%%%%%%%%%%%%%%%%%%%%%%%%%%%%%

To explain the experimental results discussed in the last section, we
have developed a theoretical model of the generation, propagation and
detection of coherent acoustic phonons in strained GaN/InGaN
heterostructures. The pump laser pulse generates a strain field that
propagates through the sample which, in turn, causes a spatio-temporal
change in the index of refraction. This change is responsible for the
oscillatory behavior seen in the probe-field reflectivity in various
semiconductor heterostructures. An approximate method of solving
Maxwell's equations in the presence of spatio-temporal disturbances in
the optical properties and obtaining the reflectivity of the probe
field in thin films excited by picosecond pump pulses can be found in
Thompsen et.~al.\cite{Tho864129}

The spatio-temporal disturbance of the refractive index is caused by
the propagating coherent phonon wavepackets. Thus, an essential
ingredient in understanding the probe reflectivity is a model for
the generation and propagation of the very short strain pulse in the
sample. Recently, a microscopic theory explaining the generation and
propagation of such strain pulses was reported by Sanders
et.~al.\cite{San01235316,San01235316a} where it was shown that
propagating coherent acoustic phonon wavepackets are created by the
nonequilibrium carriers excited by the ultrafast pump pulse. The
acoustic phonon oscillations arise through the electron-phonon
interaction with the photoexcited carriers.  Both acoustic deformation
potential and piezoelectric scattering were considered in the
microscopic model. It was found that under typical experimental
conditions, the microscopic theory could be simplified and mapped onto
a loaded string model.  Here, we use the string model of coherent
phonon pulse generation to obtain the strain field seen by the probe
pulse.

First, we solve Maxwell's equations to obtain the probe reflectivity
in the presence of a generalized spatio-temporal disturbance of the
index of refraction. Let $n_\text{b}$ be the index of refraction
without the strain which is real because initially the absorption can
be neglected and let $\delta\tilde{n}=\delta n+i\kappa$ be the
propagating change in the index of refraction due to the strain. When
the effect of the change of the index of refraction is taken into
account, the probe field with energy $\omega$ can be described by the
following generalized wave equation
\begin{equation}
\label{eewave}
\frac{\partial^2 E(z,t)}{\partial z^2} + \frac{\omega^2}{c^2}[n_\text{b}
+ \delta\tilde{n}(z,t)]^2 E(z,t) = 0
\end{equation}
where $E(z,t)$ is the probe field in the slowly varying envelope
function approximation and $\omega$ is the central frequency of the
probe pulse. Eqn.~(\ref{eewave}) is obtained from Maxwell's equations
assuming that the polarization response is instantaneous and that the
probe pulse obeys the slowly varying envelope function
approximation. Since $|\delta\tilde{n}|\ll n_\text{b}$ under typical
conditions, Eqn.~(\ref{eewave}) can be cast into
\begin{equation}
\label{eewave2}
\frac{\partial^2 E(z,t)}{\partial z^2} + (n_\text{b} k)^2 E(z,t) =
-~2n_\text{b} k^2 \delta\tilde{n}(z,t) E(z,t)
\end{equation}
where $k=\omega/c$ is the probe wavevector.  To relate the change of
the index of refraction to the strain field, $\eta(z,t)\equiv\partial
U(z,t)/\partial z$, we assume $|\delta\tilde{n}| \ll n_\text{b}$ and
adopt the linear approximation
\cite{Tho864129}
\begin{equation}
\label{edn}
\delta\tilde{n}(z,t) = \frac{\partial \tilde{n}}{\partial \eta}\eta(z,t).
\end{equation}

We view Eqn.~(\ref{eewave2}) as an inhomogeneous Helmholtz equation
and obtain the solution using the Green's function technique. The
desired Green's function is determined by solving
\begin{equation}
\label{egreen}
\frac{\partial^2 G(z,z^\prime)}{\partial z^2} + (n_\text{b} k)^2
G(z,z^\prime) = \delta(z-z^\prime)
\end{equation}
and the result is
\begin{equation}
\label{egreen-fun}
G(z,z^\prime) = -\frac{i}{2n_\text{b} k}\exp\left(in_\text{b}k
|z-z^\prime| \right).
\end{equation}
Then, the solution to Eqn.~(\ref{eewave2}) can be written as
\begin{widetext}
\begin{eqnarray}
E(z,t) &=& E_\text{h}(z,t) + \int_{-\infty}^\infty~dz^\prime
G(z,z^\prime)\left\{-2n_\text{b}k^2 \ \delta\tilde{n}(z^\prime,t)
\ E(z^\prime,t)\right\}\\
&\approx & E_\text{h}(z,t) + \int_{-\infty}^\infty~dz^\prime
G(z,z^\prime)\left\{-2n_\text{b}k^2 \ \delta\tilde{n}(z^\prime,t)
\ E_\text{h}(z^\prime,t)\right\}
\label{esol}
\end{eqnarray}
\end{widetext}
where we have chosen the lowest Born series in the last line. In
Eqn.~(\ref{esol}), $E_\text{h}$ is the homogeneous solution which
takes the form $E_\text{h}(z,t)= {\tilde E}_\text{h}(z,t)
\exp\left\{i(n_\text{b}kz-\omega t)\right\}$
where ${\tilde E}_\text{h}(z,t)$ is a slowly varying 
envelope function. This describes the probe pulse moving to the
right in the sample without optical distortion. 

We now apply this approximate solution to our structure where the
interface between air and the sample is chosen at $z=0$.  In the air,
where $z\le 0$, there is an incident probe pulse traveling toward the
sample as well as a reflected pulse. The electric field in the air,
$E_<(z,t)$, can thus be written as the sum
\begin{equation}
\label{einair}
E_<(z,t) = {\tilde E}_\text{i}(z,t) \ e^{i(kz-\omega t)} +
{\tilde E}_\text{r}(z,t) \ e^{-i(kz+\omega t)}
\end{equation}
where ${\tilde E}_\text{i}(z,t)$ and
${\tilde E}_\text{r}(z,t)$ are the slowly varying
envelope functions of the incident and reflected probe fields,
respectively. Inside the sample, $z\ge 0$, the solution is given as
\begin{widetext}
\begin{eqnarray}
\label{einside}
E_>(z,t) &= &E_\text{h}(z,t)+
\int_{0}^\infty~dz^\prime\exp\left(in_\text{b}
k(z^\prime-z)\right)ik \ \delta\tilde{n}(z^\prime,t) \ E_\text{h}(z^\prime,t)
\nonumber\\
& = & {\tilde E}_\text{t}(z,t)e^{i(n_\text{b}kz-\omega t)} +
\left\{\int_{0}^\infty dz^\prime\exp\left(2in_\text{b} kz^\prime\right)
ik \ \delta\tilde{n}(z^\prime,t)\right\}
{\tilde E}_\text{t}(z,t)e^{-i(n_\text{b}kz+\omega t)}\\
&=& {\tilde E}_\text{t}(z,t)e^{i(n_\text{b}kz-\omega t)}+
{\cal A}(n_\text{b}k,t){\tilde E}_\text{t}(z,t)e^{-i(n_\text{b}kz
+\omega t)}
\end{eqnarray}
\end{widetext}
where we used $E_\text{h}(z,t)={\tilde E}_\text{t}(z,t) \
\exp(i(n_\text{b}kz-\omega t))$ in the second step, assuming that
${\tilde E}_\text{t}(z,t)$ is nearly constant within the slowly
varying envelope function approximation, and we define a reflected
amplitude function
\begin{equation}
\label{emodul-amp}
{\cal A}(n_\text{b} k,t) \equiv \int_{0}^\infty dz^\prime \
\exp\left(2in_\text{b} k z^\prime\right)
\ ik \ \delta\tilde{n}(z^\prime,t).
\end{equation}
The expression for the reflected amplitude function, ${\cal
A}(n_\text{b} k,t)$, in Eqn.~(\ref{einside}) says there is a
frequency-dependent modulation of the amplitude in the reflected wave
in the sample due to the propagating strain.

Having determined the waves on both sides of the interface, we can now
calculate the reflectivity. We apply the usual boundary conditions to
the slowly varying envelope functions and the results are written
compactly as
\begin{equation}
\left(  \begin{array}{cc}
        -1  &  1+{\cal A} \\
        1   &  n_\text{b}(1-{\cal A})
        \end{array}
\right)
\left(  \begin{array}{c}
          {\tilde E}_\text{r} \\
          {\tilde E}_\text{t}
        \end{array}
\right) =
\left(  \begin{array}{c}
         {\tilde E}_\text{i} \\
         {\tilde E}_\text{i}
        \end{array}
\right).
\end{equation}
We solve this equation to obtain
\begin{equation}
\frac{{\tilde E}_\text{r}}{{\tilde E}_\text{i}} = \frac{r_0 + {\cal
A}}{1 + r_0{\cal A}} \approx r_0 + {\cal A}
\end{equation}
where $r_0=(1-n_\text{b})/(1+n_\text{b})$.
To the same order, we find that $\tilde E_\text{t}/\tilde E_i\approx
t_0(1-r_0{\cal A})$ where $t_0=2/(1+n_\text{b})$.
It is now straightforward to calculate the differential reflectivity as
\begin{equation}
\label{edr1}
\frac{\Delta R}{R} = \frac{|r_0 + \Delta r|^2 - |r_0|^2}
	{|r_0|^2}
\approx \frac{2}{r_0} {\rm Re}{\cal A}.
\end{equation}
Finally, by substituting the linear law Eqn.~(\ref{edn}) into
Eqn.~(\ref{emodul-amp}) and using Eqn. (\ref{edr1}) we get
\begin{equation}
\label{edr2}
\frac{\Delta R}{R} = \int_0^\infty dz \ {\cal F}(z,\omega) \
\frac{\partial U(z,t)}{\partial z}
\end{equation}
where the sensitivity function, ${\cal F}(z,\omega)$, is defined as
\begin{equation}
\label{efzw}
{\cal F}(z,\omega) = -\frac{2k}{r_0}
\left[\frac{\partial n}{\partial \eta}\sin(2n_\text{b}kz)
+\frac{\partial \kappa}{\partial \eta}\cos(2n_\text{b}kz)\right]
\end{equation}

In Eqn.~(\ref{edr2}), the differential reflectivity is expressed in
terms of the lattice displacement, $U(z,t)$, due to propagating
coherent phonons. Sanders et.~al.\cite{San01235316,San01235316a}
developed a microscopic theory showing that the coherent phonon
lattice displacement satisfies a driven string equation,
\begin{equation}
\frac{\partial^2 U(z,t)}{\partial t^2} -
C_\text{s}^2 \ \frac{\partial^2 U(z,t)}{\partial z^2} = S(z,t),
\label{estring}
\end{equation}
where $C_\text{s}$ is the LA sound speed in the medium and $S(z,t)$ is
a driving term which depends on the photogenerated carrier density.
The LA sound speed is related to the mass density, $\rho$, and the
elastic stiffness constant, $C_{33}$, by
$C_\text{s}=\sqrt{C_{33}/\rho}$.  The LA sound speed is taken to be
different in the GaN/InGaN heterostructure and Sapphire substrate.
For simplicity, we neglect the sound speed mismatch between the GaN
and In$_x$Ga$_{1-x}$N layers.

The driving function, $S(z,t)$, is nonuniform and is given by
\begin{equation}
S(z,t) = \sum_{\nu} S_{\nu}(z,t) ,
\label{eszt}
\end{equation}
where the summation index, $\nu$, runs over carrier species, i.e.,
conduction electrons, heavy holes, light holes, and crystal field
split holes, that are created by the pump pulse. Each carrier species
makes a separate contribution to the driving function.  The partial
driving functions, $S_{\nu}(z,t)$, in piezoelectric wurtzite crystals
depend on the density of the photoexcited carriers. Thus,
\begin{equation}
S_{\nu}(z,t) = \pm \frac{1}{\rho} \left\{
a_{\nu} \frac{\partial}{\partial z}+\frac{4 \pi |e| \
e_{33}}{\epsilon_{\infty}}
\right\} \rho_{\nu}(z,t) ,
\label{esnu}
\end{equation}
where the plus sign is used for conduction electrons and the minus
sign is used for holes. Here $\rho_{\nu}(z,t)$ is the photogenerated
electron or hole number density, which is real and positive, $\rho$ is
the mass density, $a_{\nu}$ are the deformation potentials for
each carrier species, $e_{33}$
is the piezoelectric constant, and $\epsilon_{\infty}$ is the high
frequency dielectric constant.

To be more specific, we will consider a SQW sample of the type shown
in Fig.~\ref{fsamples}(c).  We adopt a simple model for
photogeneration of electrons and holes in which the photogenerated
electron and hole number densities are proportional to the squared
ground state particle in a box wavefunctions. The exact shape of the
electron and hole number density profile is not critical in the
present calculation since all that really matters is that the
electrons and holes be strongly localized. The carriers are assumed to
be instantaneously generated by the pump pulse and are localized in
the In$_x$Ga$_{1-x}$N quantum well. Having obtained a model expression
for $\rho_{\nu}(z,t)$, it is straightforward to obtain $S(z,t)$ using
Eqns.~(\ref{eszt}) and (\ref{esnu}).

To obtain $U(z,t)$, we solve driven string equations in the GaN
epilayer and the Sapphire substrate, namely
\begin{subequations}
\label{WaveEquations}
\begin{equation}
\frac{\partial^2 \ U(z,t)}{\partial t^2} -
C_0^2 \ \frac{\partial^2 \ U(z,t)}{\partial z^2} = S(z,t)
\ \ \ \ \ \ \ (0 \le z \le L)
\label{GaNWave}
\end{equation}
and
\begin{equation}
\frac{\partial^2 \ U(z,t)}{\partial t^2} -
C_1^2 \ \frac{\partial^2 \ U(z,t)}{\partial z^2} = 0
\ \ \ \ \ \ \ \ \ \ \ \ \ (L \le z \le Z_\text{s})
\label{SapphireWave}
\end{equation}
\end{subequations}
where $C_0$ and $C_1$ are LA sound speeds in the GaN and Sapphire
substrate, respectively. In Eqn.~(\ref{SapphireWave}), the Sapphire
substrate has finite thickness. To simulate coherent phonon
propagation in an infinite Sapphire substrate, $Z_\text{s}$ in
Eqn.~(\ref{SapphireWave}) is chosen large enough so that the
propagating sound pulse generated in the GaN epilayer does not have
sufficient time to reach $z = Z_\text{s}$ during the simulation. If
$T_{sim}$ is the duration of the simulation, this implies $Z_\text{s}
\ \ge \ L + C_1 \ T_{sim}$.

Equations (\ref{GaNWave}) and (\ref{SapphireWave}) are solved subject
to initial and boundary conditions. The initial conditions are
\begin{equation}
U(z,0) = \frac{\partial U(z,0)}{\partial t} = 0 .
\label{InitialConditions}
\end{equation}
At the GaN-air interface at $z = 0$, we assume the free surface
boundary condition
\begin{subequations}
\label{BoundaryConditions}
\begin{equation}
\frac{\partial U(0,t)}{\partial z} = 0
\end{equation}
since the air exerts no force on the GaN epilayer.
The phonon displacement and the force per unit area are continuous at
the GaN-Sapphire interface so that
\begin{equation}
U(L-\epsilon,t) = U(L+\epsilon,t)
\end{equation}
and
\begin{equation}
\rho_0 \ C_0^2 \ \frac{\partial U(L-\epsilon,t)}{\partial z} =
\rho_1 \ C_1^2 \ \frac{\partial U(L+\epsilon,t)}{\partial z}.
\end{equation}
The boundary condition at $z=Z_\text{s}$ can be chosen arbitrarily
since the propagating sound pulse never reaches this interface. For
mathematical convenience, we choose the rigid boundary condition
\begin{equation}
U(Z_\text{s},t) = 0.
\end{equation}
\end{subequations}

To obtain $U(z,t)$ for general $S(z,t)$, we first need to find the
harmonic solutions in the absence of strain, i.e. $S(z,t) = 0$. The
harmonic solutions are taken to be
\begin{equation}
U_n(z,t) =
W_n(z) \ e^{-i \omega_n t} \ \ \ \ \ \ ( \ \omega_n \ge 0 \ )
\end{equation}
and it is easy to show that the mode functions, $W_n(z)$, satisfy
\begin{subequations}
\begin{equation}
\frac{d^2 W_n(z)}{dz^2} + \frac{\omega_n^2}{C_0^2} W_n(z) = 0
\ \ \ \ \ \ ( \ 0 \le z \le L \ )
\end{equation}
and
\begin{equation}
\frac{d^2 W_n(z)}{dz^2} + \frac{\omega_n^2}{C_1^2} W_n(z) = 0
\ \ \ \ \ \ ( \ L \le z \le Z_\text{s} \ )
\end{equation}
\end{subequations}
Applying the boundary conditions (\ref{BoundaryConditions}) we obtain
the mode functions
\begin{subequations}
\begin{equation}
W_n(z) =
\left\{
\begin{array}{ll}
\cos \left( \omega_n z/C_0 \right) &
\ \ \mbox{if \ $0 \le z \le L$}
\\
B_n \ \sin \left( \omega_n (Z_\text{s}-z)/C_1 \right) &
\ \ \mbox{if \ $L \le z \le Z_\text{s}$}
\end{array}
\right.
\end{equation}
with
\begin{equation}
B_n=\frac{\cos \left( \omega_n L/C_0     \right)}
         {\sin \left( \omega_n (Z_\text{s}-L)/C_1 \right)}.
\end{equation}
\end{subequations}
The mode frequencies, $\omega_n$, are solutions of the transcendental
equation
\begin{equation}
\frac{1}{\rho_0 C_0}\cot\left( \frac{\omega_n L}{C_0} \right) =
\frac{1}{\rho_1 C_1}\tan\left( \frac{\omega_n (Z_\text{s}-L)}{C_1}
\right)
\end{equation}
which we solve numerically to obtain the mode frequencies,
$\omega_n$ ($n = 0, 1, 2, ...$). The index, $n$, is
equal to the number of nodes in the mode functions, $W_n(z)$.

A general displacement can be expanded in terms of the harmonic modes
as
\begin{equation}
U(z,t)=\sum_{n=0}^{\infty} \ q_n(t) \ W_n(z).
\label{USeries}
\end{equation}
Substituting equation (\ref{USeries}) for $U(z,t)$ into
(\ref{WaveEquations}) and taking the initial conditions
(\ref{InitialConditions}) into account, we find that the expansion
coefficients, $q_n(t)$, satisfy a driven harmonic oscillator equation
\begin{equation}
\ddot{q}_n(t) + \omega_n^2 \ \dot{q}_n(t) = Q_n(t),
\end{equation}
subject to the initial conditions $q_n(0)=\dot{q}_n(0)=0$.
The harmonic oscillator driving term $Q_n(t)$ is given by
\begin{equation}
Q_n(t)=\frac{\int_0^{Z_\text{s}} dz \ W_n(z)\
S(z,t)}{\int_0^{Z_\text{s}} dz \
W_n(z)^2}.
\end{equation}

In our simple displacive model for photogeneration of carriers,
$S(z,t)=S(z) \ \Theta(t)$ where $\Theta(t)$ is the Heaviside step
function. In this case, the lattice displacement is explicitly given
by
\begin{equation}
U(z,t)=\sum_{n=0}^{\infty}
\frac{S_n}{\omega_n^2} ( \ 1-\cos(\omega_n \ t) \ ) \ W_n(z)
\label{Uzt}
\end{equation}
with $S_n$ defined as
\begin{equation}
S_n=\frac{\int_0^{Z_\text{s}} dz \ W_n(z)\
S(z)}{\int_0^{Z_\text{s}} dz \ W_n(z)^2}.
\end{equation}
Using the lattice displacement (\ref{Uzt}), we obtain the
time-dependent differential reflectivity at the probe frequency,
$\omega$, from equation (\ref{edr2}). The result is
\begin{equation}
\frac{\Delta R}{R}(\omega,t)=\sum_{n=0}^{\infty}
\frac{S_n}{\omega_n^2} ( \ 1-\cos(\omega_n \ t) \ ) \ R_n(\omega)
\label{Rzt}
\end{equation}
where
\begin{equation}
R_n(\omega) =
\int_0^{Z_\text{s}} dz \ {\cal F}(z,\omega) \ \frac{dW_n(z)}{dz}.
\end{equation}
can be evaluated analytically.

With the above formalism, we solve for the lattice displacement,
$U(z,t)$, for a coherent LA phonon pulse propagating in a multilayer
structure consisting of a 1.124~$\mu$m thick GaN epilayer grown on top
of an infinitely thick Sapphire substrate with the growth direction
along $z$. We take the origin to be at the GaN-air interface and the
GaN-Sapphire interface is taken to be at $z=L=1.124$~$\mu$m. We assume
that carriers are photogenerated in a single 240~\AA\ thick
In$_{x}$Ga$_{1-x}$N quantum well embedded in the GaN layer 0.1~$\mu$m
below the GaN-air interface and 1~$\mu$m above the Sapphire substrate.
Our structure thus resembles the SQW sample shown in
Fig.~\ref{fsamples}(c).  In the GaN epilayer, we take $C_{33}=379$~GPa
and $\rho_0=6.139$~gm/cm$^3$ \ \cite{Amb993222} from which we obtain
$C_0=7857$~m/s.  For the Sapphire substrate, we take $C_{33}= 500$~GPa
and $\rho_1=3.986$~gm/cm$^3$ \ \cite{Eve842190} from which we find
$C_1=11200$~m/s.

%...................................................................
%                               figure 7
%...................................................................
\begin{figure}[tbp]
\begin{center}
\includegraphics[width=3.4in]{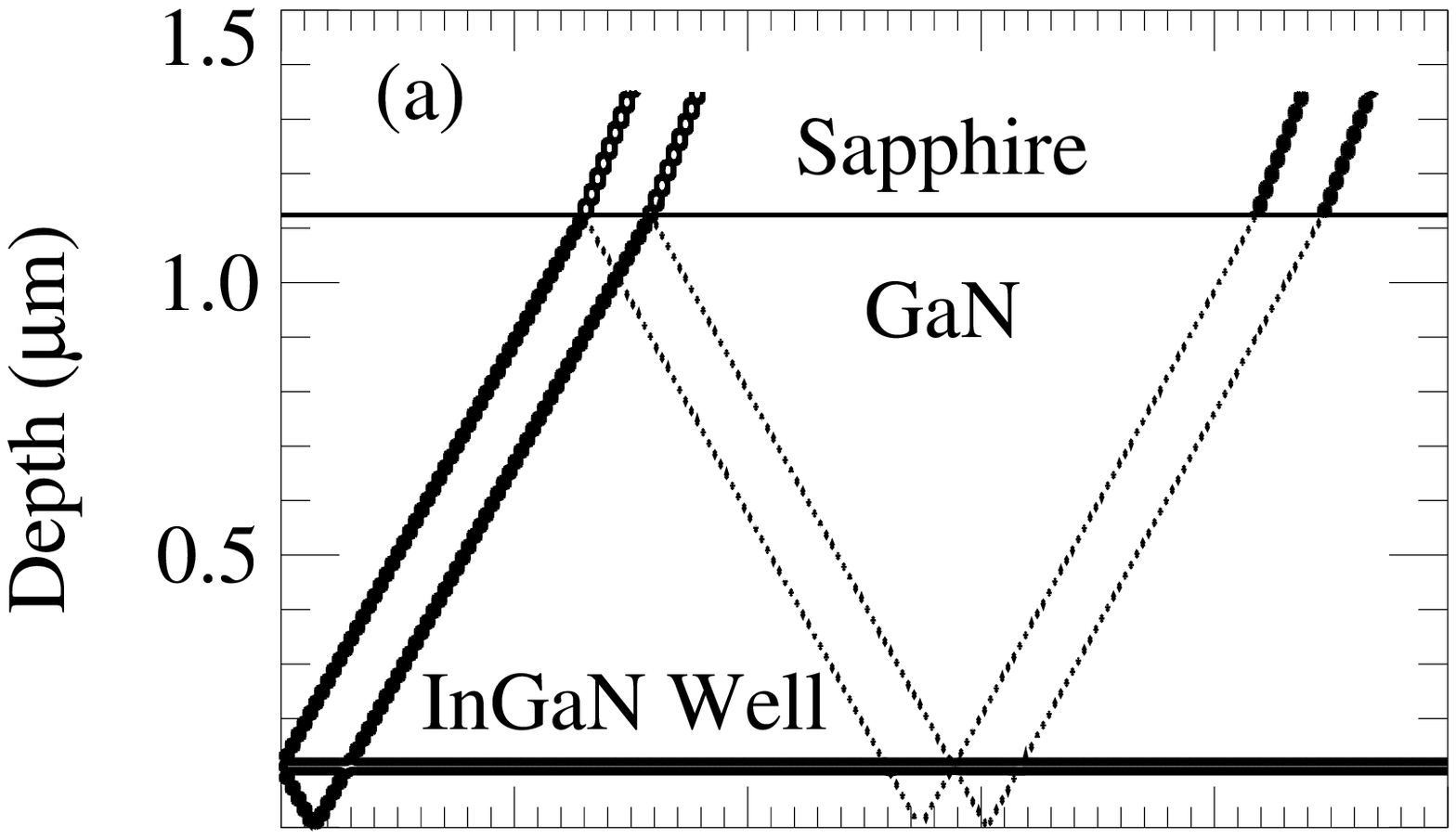}
\vskip -0.45in
\includegraphics[width=3.4in]{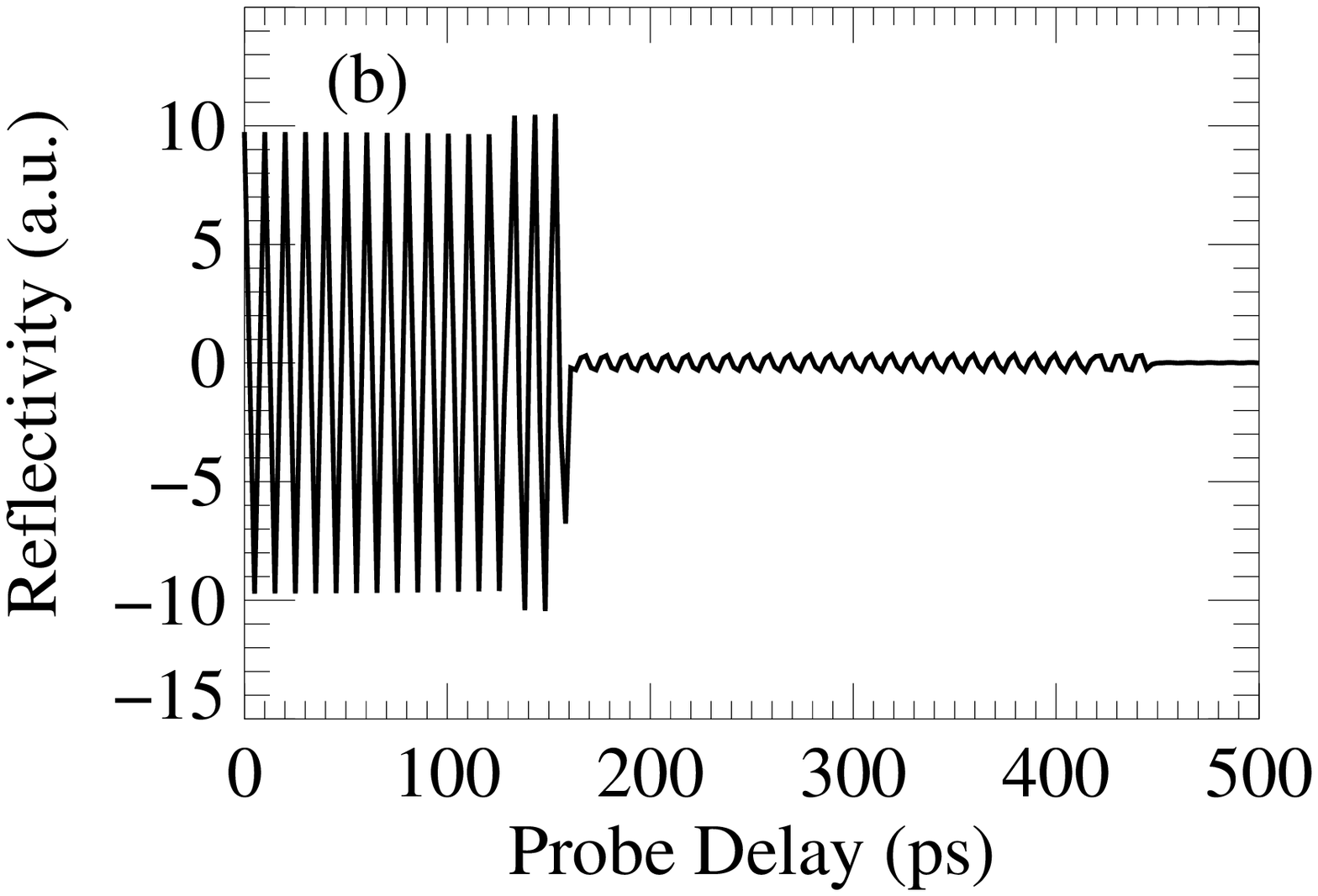}
\end{center}
\caption{
Generation and propagation of coherent acoustic phonons photogenerated
in a single InGaN well embedded in a free standing $1.124 \mu$m GaN
epilayer grown on top of a Sapphire substrate.  In (a) a contour plot
of the strain field, $\partial U(z,t)/\partial z$, is shown as a
function of depth below the GaN-air interface and the probe delay.  In
(b), the resulting differential reflectivity induced by the strain
field in (a) is shown as a function of the probe delay.  }
\label{fsanders}
\end{figure}
%..................................................................

The results of our simulation are shown in Fig.~\ref{fsanders}.  A
contour map of the strain, $\partial U(z,t)/\partial z$, is shown in
Fig.~\ref{fsanders}(a). We plot the strain as a function of the depth
below the surface and the probe delay time. Photoexcitation of
electrons and holes in the InGaN quantum well generates two coherent
LA sound pulses traveling in opposite directions. The pulses are
totally reflected off the GaN/air interface at $z=0$ and are partially
reflected at the Sapphire substrate at $z=1.124 \ \mu$m.
Approximately 95\% of the pulse energy is transmitted and only 5\% is
reflected at the substrate. The speed of the LA phonon pulses is just
the slope of the propagating wave trains seen in
Fig.~\ref{fsanders}(a) and one can clearly see that the LA sound speed
is greater in the Sapphire substrate.

From the strain the differential reflectivity can be obtained from
equation (\ref{edr2}). From Fig.~\ref{fsanders}(a), the strains,
$\partial U(z,t)/\partial z$, associated with the propagating pulses
are highly localized and travel at the LA sound speed. Each pulse
contributes a term to the differential reflectivity that goes like
\begin{equation}
\frac{\Delta R}{R}(\omega,t) \sim {\cal F}(C_0 t,\omega) \propto
\frac{\omega}{c} \sin \left( \frac{2 n_\text{b} \omega}{c} C_0 t
+ \phi \right)
\label{DRmicro}
\end{equation}
The period of the oscillations of ${\cal F}$ depends on the probe
wavelength, $\lambda = 2 \pi c/\omega$, with the result that the
observed differential reflectivity oscillates in time with period, $T
= \pi c /(n_\text{b} C_0 \omega) = \lambda/(2 n_\text{b} C_0)$, where
$n_\text{b}=2.4$ is the index of refraction, and $C_0=7857$ m/s is the
LA sound speed in GaN.  For $\lambda = 377$ nm ($\hbar \omega = 3.29$
eV), this gives us $T=10$ ps. The sensitivity function, ${\cal
F}(z,\omega)$, defined in equation (\ref{efzw}) is an oscillating
function in the GaN/InGaN epilayer and is assumed to vanish in the
Sapphire substrate. Our computed differential reflectivity is shown in
Fig.~\ref{fsanders}(b) for a probe wavelength of $\lambda = 377 \
\mbox{nm}$.  We find that the reflectivity abruptly attenuates when
the strain pulse enters the Sapphire substrate at $t=170$ ps. The
reflected strain pulses give rise to the weaker oscillations seen for
$t > 170$ ps. These oscillations are predicted to continue until the
reflected pulses are again partially reflected off the Sapphire
substrate at $t = 430$ ps.

%%%%%%%%%%%%%%%%%%%%%%%%%%%%%%%%%%%%%%%%%%%%%%%%%%%%%%%%%%%%%%%%%%%%
\section{Simple model}
%%%%%%%%%%%%%%%%%%%%%%%%%%%%%%%%%%%%%%%%%%%%%%%%%%%%%%%%%%%%%%%%%%%%

Since the coherent oscillation observed in the differential
reflectivity stems essentially from the strain pulse propagating into
the layers, most phenomena can be understood by a simple macroscopic
model that is presented in this section.

%...................................................................
%                            figure 8
%...................................................................
\begin{figure}[tbp]
\begin{center}
\includegraphics[width=3.4in]{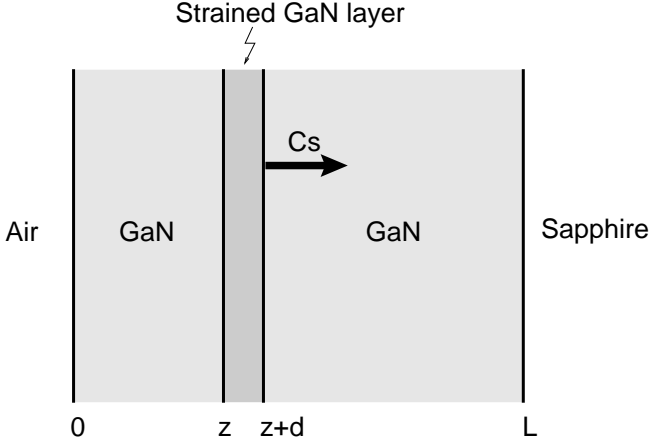}
\end{center}
\caption{Propagating strained GaN layer in our
simple model. The pump laser pulse creates a coherent acoustic phonon
wavepacket in the InGaN layer near the air/GaN surface, which is
modelled as a thin strained layer. The strained GaN layer propagates
into the host GaN layer. The index of refraction in the strained layer
is perturbed relative to the background GaN due to the strong
strain induced piezoelectric field (Franz-Keldysh effect).}
\label{fmodel}
\end{figure}
%...................................................................

Instead of solving the loaded string equations for the strain to
obtain a propagating disturbance in the refractive index, the
propagating strain pulse at a given moment can be viewed as a thin
strained layer in the sample, where the index of refraction is assumed
to be slightly different from the rest of the sample. This situation
is schematically depicted in Fig.~\ref{fmodel} where a fictitious,
thin GaN strained layer is located at $z$ in the thick host GaN
layer. The thickness of the strained layer, $d$, is approximately the
width of the traveling coherent phonon strain field, $\partial
U(z,t)/\partial z$, and is to be determined from the microscopic
theory. From the last section, it was seen that the propagating strain
field is strongly localized so that $d$ is small. In the example of
the last section, $d$ is approximately equal to the quantum well
width.  Here we assume the strain pulse has been already created near
the air/GaN interface and do not consider its generation procedure.
We treat $d$ as a phenomenological constant and also assume that the
change in the index of refraction is constant. This strained GaN layer
travels into the structure with the speed of the acoustic phonon
wavepacket $C_0=7\times 10^3~m/s$, so the location of the stained
layer is given as $z=C_0\tau$ where $\tau$ is the pump-probe delay
time.

Within the slowly varying envelope function approximation, the
solutions to the Maxwell equation can be written as plane
waves in each region,
\begin{equation}
\label{eEB}
E_i(z,t)=a_ie^{ik_iz_i-i\omega t}+b_ie^{-ik_iz_i-i\omega t}
\end{equation}
where $E_i$ is the electric field in the left layer of $i$th
interface, and $a_i$ and $b_i$ are the slowly varying amplitudes.  The
magnetic field is given by $B_i\sim \partial{E_i}/\partial z$.  For a
normally incident probe field, we apply the usual matching conditions
on $E$ and $B$ to obtain
\begin{equation}
\label{eitoi1}
\left( \begin{array}{c} a_i \\ b_i
\end{array} \right) = {\mathcal M}_i
\left( \begin{array}{c} a_{i+1} \\ b_{i+1}
\end{array} \right),
\end{equation}
where the transfer matrix, \({\mathcal M}_i\), is given explicitly as
\begin{eqnarray}
\label{emi}
&&{\mathcal M}_i =\\
&&\frac{1}{2}\left( \begin{array}{cc}
(1+\frac{k_{i+1}}{k_i})e^{i(k_{i+1}-k_i)z_i} & 
(1-\frac{k_{i+1}}{k_i})e^{-i(k_{i+1}+k_i)z_i} \\
(1-\frac{k_{i+1}}{k_i})e^{i(k_{i+1}+k_i)z_i} &
(1+\frac{k_{i+1}}{k_i})e^{-i(k_{i+1}-k_i)z_i}\nonumber
\end{array} \right).
\end{eqnarray}
To apply this formula to our configuration, we normalize the incident
amplitude to $1$, let $r$ be the reflected amplitude in the air, and
impose the boundary condition that there is only a transmitted wave
into the Sapphire substrate with amplitude $t$ and no reflected wave
from the GaN-Sapphire interface back into the GaN epilayer. The latter
assumption is reasonable since the microscopic theory of the previous
section shows that only 5\% of the pulse energy is reflected from the
interface between the GaN and the Sapphire substrate. The total
reflection and transmission amplitudes $r$ and $t$ for the GaN
epilayer structure are determined by
\begin{equation}
\label{emt}
\left( \begin{array}{c} 1 \\ r
\end{array}\right)
= \left( \begin{array}{cc} {\cal M}_{11} & {\cal M}_{12} \\
{\cal M}_{21} & {\cal M}_{22}
\end{array}\right)
\left( \begin{array}{c} t \\ 0
\end{array}\right)
\end{equation}
where ${\cal M}={\mathcal M}_1{\mathcal M}_2\cdots{\mathcal M}_n$.
The reflection amplitude is readily found to be
\begin{equation}
r={\cal M}_{21}/{\cal M}_{11}=r_0+\Delta r \label{ert},
\end{equation}
where $r_0$ is the background contribution without the strained
layer. The total transmission amplitude is given as $t=1/{\cal
M}_{11}$.  We can now numerically calculate the differential
reflectivity by substituting $r$ into Eqn.~(\ref{edr1}).

%...................................................................
%                            figure 9
%...................................................................
\begin{figure}[tbp]
\begin{center}
\includegraphics[width=3.4in]{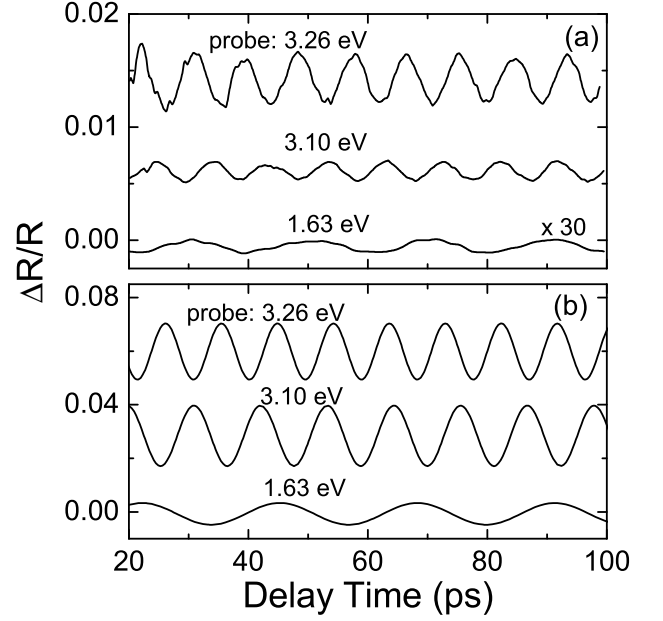}
\end{center}
\caption
{Differential reflection for different frequencies of the
probe pulse. Plot (a) is experimental data and (b) comes from our calculations described in the text.}
\label{frtwcmp}
\end{figure}
%..................................................................

In Fig.~\ref{frtwcmp} we compare our theoretical model with
experimental results. In Fig.~\ref{frtwcmp}(a), the oscillatory part
of the differential reflectivity is plotted as a function of probe
delay for three values of the probe photon energy.  This is the same
data that was shown in Fig.~\ref{fxrtw} where it was seen that the
oscillation period of the coherent phonon reflectivity oscillations
are proportional to the probe wavelength. In Fig.~\ref{frtwcmp}(b), we
plot the corresponding theoretical differential reflectivities
obtained from Eqn.~(\ref{edr1}).  Comparing Figs.~\ref{frtwcmp}(a) and
\ref{frtwcmp}(b), we see that our geometrical optics model
successfully explains the observed relation between the coherent
phonon oscillation period and the probe wavelength. However, the
calculated amplitudes of oscillation is inconsistent with the
experimental data because in this case we have not taken into account
of the change of index of refraction with respect to the probe
wavelength which we will do later in the paper.

%...................................................................
%                             figure 10
%...................................................................
\begin{figure}[tbp]
\begin{center}
\includegraphics[width=3.4in]{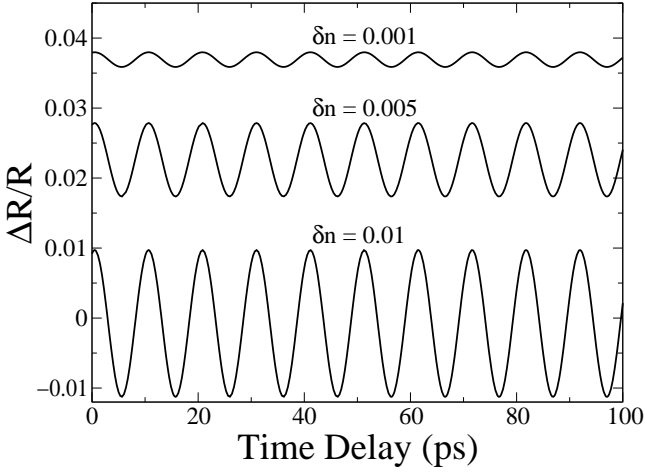}
\end{center}
\caption{The differential reflection calculated numerically using
\protect{Eqn.~(\ref{edr1})} for three different values in the
change in index of refraction. The parameters used are
$n_\text{b}=2.65$,
$\omega=3.29~eV$, $d=30~nm$, $L=1~\mu m$, and $C_\text{s}=7000~m/s$.
[Note that the curves are sifted to avoid overlapping.]}
\label{frtn}
\end{figure}
%...................................................................

\begin{figure}[tbp]
\begin{center}
\includegraphics[width=3.4in]{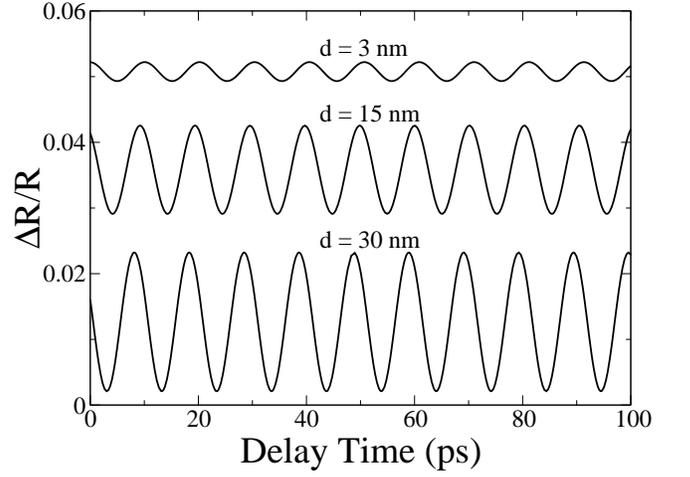}
\end{center}
\caption{Calculated differential reflection for different values of
the thickness of the strained GaN layer. The parameters are
$n_\text{b}=2.65$, $\omega=3.29~eV$, $L=1~\mu m$, and
$C_\text{s}=7000~m/s$.  The change in the index of refraction
in the strained GaN layer was assumed to be $\delta n=0.01$.}
\label{frtd}
\end{figure}

In Fig.~\ref{frtn} we plot the differential reflectivity as a function
of the probe delay for three different values of the change in the
index of refraction in the strained layer, $\delta\tilde{n}$.  As
expected, the greater the change in index of refraction, the greater
the amplitude of the differential reflectivity oscillations.  A larger
change in the index of refraction implies that more electron-hole
pairs are excited near the air-GaN interface, which acts as a stronger
source for the coherent acoustic phonon reflectivity
oscillations. These results are qualitatively consistent with the
experimental results shown in Fig.~\ref{fxrtx}.  In Fig.~\ref{frtd} we
fix the change in the index of refraction, $\delta n=0.01$, and varied
the thickness of the strained layer, $d$.  The result shows a larger
amplitude for the differential reflection in wider strained layers,
which is consistent with what was experimentally observed in
Fig.~\ref{dqw3}.

%...................................................................
%                             figure 11
%...................................................................
%...................................................................

\begin{figure}[tbp]
\begin{center}
\includegraphics[width=3.4in]{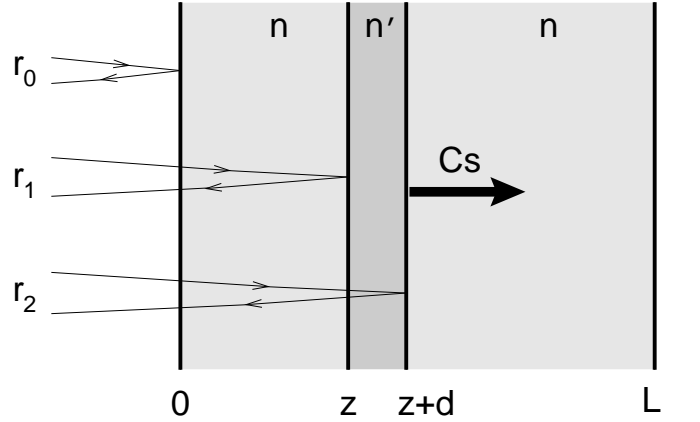}
\end{center}
\caption{Schematic diagram of the single-reflection approximation
in the Fabry-Perot reflection; where we have selected the processes
only proportional to $\delta\tilde{n}$ ($=n^\prime-n$).}
\label{fr012}
\end{figure}

Both Fig.~\ref{frtn} and Fig.~\ref{frtd} suggest that the amplitude of
the differential reflectivity seems to increase monotonically with
$\delta n$ and $d$.  All these features can be understood more easily
using the single-reflection approximation to the full formula for the
reflectivity in Eqn. (\ref{ert}). Assuming the change in the index of
refraction is small, $|\delta\tilde{n}|\ll 1$, we may select
contributions from only terms proportional to $\delta\tilde{n}$ in the
infinite Fabry-Perot series for the total reflection amplitude. The
relevant reflection processes selected are schematically depicted in
Fig.~\ref{fr012}.  In this case the total reflection amplitude is
given by the leading terms in the Fabry-Perot series
\begin{equation}
\label{er012}
r = r_0 + r_1 + r_2 +{\cal O}\left(\delta\tilde{n}^2\right)
\end{equation}
where $r_0$ is the background reflection amplitude, and the first order
terms in $\delta\tilde{n}$ are
\begin{equation}
r_1 \propto -~e^{2ikz}\delta\tilde{n},\quad r_2 \propto
e^{i2k(z+d)}\delta\tilde{n}.
\end{equation}
%

%...................................................................
%                            figure 12
%...................................................................
%...................................................................

To linear order in $\delta\tilde{n}$, the differential reflectivity
Eqn.~(\ref{edr1}) becomes
\begin{eqnarray}
\label{edr3}
\frac{\Delta R}{R}
&=&\frac{8\sin(kd)}{n^2-1}[\delta n\sin(2kz+kd)
+\kappa\cos(2kz+kd)] \nonumber \\
&=&\frac{8\sin(kd)}{n^2-1}|\delta\tilde{n}|\sin(kd)\sin(2kz+\phi),
\end{eqnarray}
where the phase $\phi$ is given by
\begin{equation}
\phi=kd + \arctan\left(\frac{\kappa}{\delta n}\right).
\end{equation}
Note that for small values of $kd$, we can expand
Eq.~(\ref{edr3}) in a Taylor series to obtain
\begin{equation}
\frac{\Delta R}{R}=\frac{8 \ kd}{n^2-1}
\left[\delta n\sin(2kz)+\kappa\cos(2kz)\right]
\label{drrkd}
\end{equation}
to first order in $kd$.

From Eqn.~(\ref{edr3}) one can see that the amplitude of the
oscillation scales linearly with the change of the real part and
imaginary part of the index of refraction respectively.

Since the strain-pulse-front moves with the speed of sound,
$z=C_0 t$, one can rewrite Eqn.~(\ref{edr3}) as
\begin{equation}
\label{edr4}
\frac{\Delta R}{R}(\omega,t) \propto |\delta\tilde{n}|
\sin(kd)\sin\left(2\pi\frac{t}{T} +\phi\right)
\end{equation}
where $T=\pi c/(n_\text{b} C_0 \omega)=\lambda/(2 C_0 n_\text{b})$.
Note that in the limit $kd \rightarrow 0$, Eqn.~(\ref{edr4})
is equivalent to Eqn.~(\ref{DRmicro}) of the microscopic theory
described in the last section.

Equation~(\ref{edr4}) shows that the period of oscillations in the
differential reflectivity is given by $T$ and hence is determined by
the wavelength of the probe pulse.  This is consistent with what was
seen in these experiments, but differs from previous experiments where
the frequency of coherent phonon oscillation was determined by the
sample and not the external probe.  Note also that Eq.~(\ref{edr4})
explains the experimental dependence of the the amplitude of the
oscillation.  It is proportional to the change in the index of
refraction and varies with $\sin(kd)$, and hence will increase
linearly with the thickness of the strained layer as well as the probe
frequency $\omega$ for $kd=n_\text{b}(\omega/c)d \ll 1$.

\begin{figure}[htbp]
\begin{center}
\includegraphics[width=3.4in]{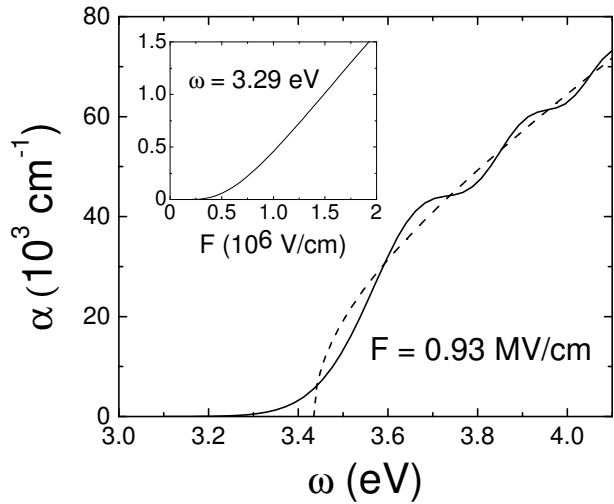}
\end{center}
\caption{The absorption coefficient as a function of the probe
energy.  The absorption tail below the band gap $E_g=3.43$~eV is
due to the built-in Franz-Keldysh field.  The dotted curve is
the free-carrier absorption.  The inset shows the absorption
coefficient as a function of the piezoelectric field.  The
parameters used in this model calculation are: $d$=30~nm,
$n_\textit{b}$=2.65, $a_0$=4~nm, and $E_0$=15~meV.}
\label{fawf}
\end{figure}

The change of index of refraction in eqn.~(\ref{edr4}) comes from two
effects. The first one, which is small, is the band gap change
(renormalization) due to the deformational potential phonon
interaction.  The other effect results from the change in the built-in
piezoelectric field which leads to non-zero absorption below the band
gap because of the Franz-Keldysh effect as shown in figure~\ref{fawf}.

In general, it is difficult to determine the Franz-Keldysh field
experimentally. Here we provide a way to estimate the
order-of-magnitude of the built-in piezoelectric field from the
strength of the measured differential reflectivity.  In
Fig.~\ref{fawf} we display the bulk absorption coefficient as a
function of photon energy, $\hbar \omega$, in bulk GaN (dotted line)
at a fixed piezoelectric field, $F = 0.93~ MV/cm$ (solid line).  The
inset shows the absorption coefficient at a fixed photon energy $\hbar
\omega = 3.29 \ eV$ as a function of the piezoelectric field, $F$. The
change in the absorption at $\hbar \omega=3.29 ~eV$ (below the
band-gap $E_g=3.43~eV$), can be used to estimate the piezoelectric
field.  For a probe energy below band-gap, we read off the amplitude
of the differential reflectivity from the experimental data. From this
value, and an estimate for $d$ based on information about the sample
geometry, we obtain the change in the index of refraction which, in
turn, gives an estimate for the absorption coefficient
$\alpha(\omega)$ at the probe photon energy.  In the Franz-Keldysh
effect, the electro-absorption $\alpha(\omega)$ is related to the
piezoelectric field, $F$, via \cite{Hau90QT}
\begin{equation}\label{Frantz}
\alpha(\omega) \simeq \frac{1}{n_\text{b}}\frac{\omega}{c}
\frac{f}{E_g-\hbar\omega} \exp{\left\{-\frac{4}{3f}
\left(\frac{(E_g-\hbar\omega)}{E_0a_0^2}\right)^{3/2} \right\}}.
\end{equation} 
Here $f=eF/(E_0a_0^2)$ where $E_0=3.435 \ \mbox{eV}$ and $a_0=39.68$
\AA\ are the excitonic energy and length scales in GaN.  From this
formula, for instance, we estimate that a piezoelectric field on the
order of $F\approx 0.93$~MeV/cm is responsible for a 1$\%$ change in
the index of refraction near the band-edge.

%...................................................................
%                             figure 13
%...................................................................
%...................................................................

%
%...................................................................
%                          figure 14
%...................................................................
\begin{figure}[!htbp]
\begin{center}
\includegraphics[width=3.4in]{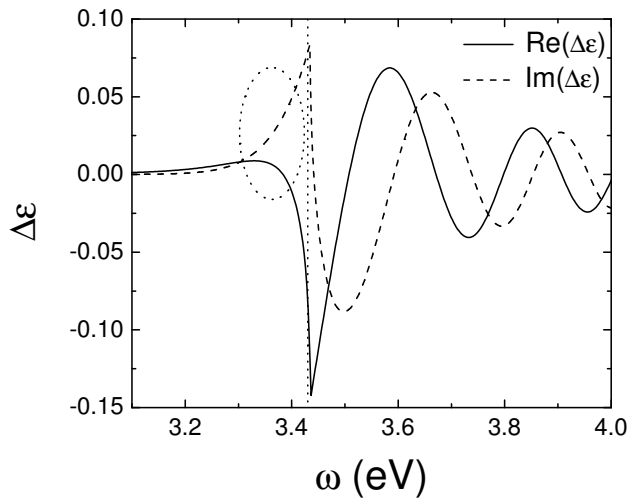}
\end{center}
\caption{The change in the real and imaginary parts of the dielectric
function in GaN as a function of the probe energy for the situation
described in Fig.~\ref{fawf}. The vertical dotted line at 3.43 eV
is the band gap energy. Just below the band gap the change in 
the dielectric function is dominated by the imaginary part as
indicated by the dotted oval.}
\label{fdielec}
\end{figure}
%...................................................................

The corresponding changes in the real and imaginary parts of the
dielectric function are shown in Fig.~\ref{fdielec}.

Finally, we would like to mention the possibility of using multiple
pump pulses to coherently control the generated coherent phonons in
the epilayers.  There have already been quite a few coherent control
experiments with multiple quantum wells from several different groups
such as Sun~\cite{Sun011201}, \"Ozg\"ur~\cite{Ozg015604}, and
Nelson~\cite{Nel03374}. In the InGaN/GaN epilayer systems, the effect
of multiple pump pulses should be more prominent then in AlGaAs/GaAs
systems because of the stronger piezoelectric field.  This might lead
to novel applications in imaging or possible new devices based on
coherent acoustic phonons.

%%%%%%%%%%%%%%%%%%%%%%%%%%%%%%%%%%%%%%%%%%%%%%%%%%%%%%%%%%%%%%%%%%%%
\section{Conclusion}
%%%%%%%%%%%%%%%%%%%%%%%%%%%%%%%%%%%%%%%%%%%%%%%%%%%%%%%%%%%%%%%%%%%%
We have presented a theory for the detection of a new class of large
amplitude, coherent acoustic phonon wavepackets in femtosecond
pump-probe experiments on In$_x$Ga$_{1-x}$N/GaN epilayers and
heterostructures. The InGaN/GaN structures are highly strained and at
high In concentrations have large built in piezoelectric fields which
account for the large amplitude of the observed reflectivity
oscillations. This new class of coherent acoustic phonons is generated
near the surface and propagates into the structure. The frequency of
the reflectivity oscillations is found to be proportional to the
frequency of the probe. These coherent phonon wavepackets can be used
as a powerful probe of the structure of the sample.  We are able to
model the generation and propagation of these acoustic phonon
wavepackets using a simple string model which is derived from a
microscopic model for the photogeneration and propagation of coherent
acoustic phonon wavepackets in InGaN/GaN multiple quantum wells. Our
model successfully predicts the observed dependence of the coherent
phonon reflectivity oscillations on probe wavelength and epilayer
thickness.

%%%%%%%%%%%%%%%%%%%%%%%%%%%%%%%%%%%%%%%%%%%%%%%%%%%%%%%%%%%%%%%%%%%%%%%
%                        Acknowledgments
%%%%%%%%%%%%%%%%%%%%%%%%%%%%%%%%%%%%%%%%%%%%%%%%%%%%%%%%%%%%%%%%%%%%%%%

\begin{acknowledgments}
This work was supported by the National Science Foundation through
grant DMR 9817828. CSK acknowledges the support from the Korea Research
Foundation. We gratefully acknowledge useful discussions with
Chi-Kuang Sun.
\end{acknowledgments}

\bibliography{paper}
\end{document}